\documentstyle[preprint,aps,epsfig]{revtex}
\begin{document}
\draft
\bigskip
\title{
Color-octet mechanism and \\ $J/\psi$ polarization at LEP 
} 
\author{Seungwon Baek\thanks{ swbaek@phya.snu.ac.kr}$^{(a)}$, 
P. Ko\thanks{ pko@phyc.snu.ac.kr}$^{(b)}$ ,
Jungil Lee\thanks{ jungil@fire.snu.ac.kr}$^{(a)}$, and 
H.S. Song\thanks{ hssong@physs.snu.ac.kr}$^{(a)} $}
\address{$^{(a)}$ 
Center for Theoretical Physics and
Department of Physics,\\ Seoul National University, 
Seoul 151-742, Korea \\
$^{(b)}$ 
Department of Physics, Hong-Ik University, Seoul 121-791, Korea
}
\date{\today}
\maketitle
\begin{abstract}
Polarized heavy quarkonium productions in  $Z^0$ decays are 
considered. We find that polarizations of the produced
quarkonia are independent of that of the 
parent $Z^0$ provided that
one  considers the energy distribution or the total production rate.
Produced $J/\psi$'s via the color-octet and the color-singlet mechanisms 
are expected to be 19\% and 29\% longitudinally polarized, respectively.
The energy dependence of 
$\eta_{1,8}(x) \equiv \frac{d\Gamma_{1,8}^L}{d x}/ 
\frac{d\Gamma_{1,8}}{d x}$ is very sensitive to the production mechanism,
and therefore the measurement of  $\eta(x)_{\rm exp}$ will be an independent
probe of the color-octet mechanism.
\end{abstract}


\narrowtext

Since Braaten and Fleming put forward the idea of the color-octet mechanism 
\cite{fleming} as a possible solution to the so-called $\psi^{'}$ puzzle 
at the Tevatron  \cite{mangano}, there have been 
many activities applying this idea to other processes : heavy
quarkonium (both $S-$ and $P-$wave charmonium and bottomium) 
productions at  the Tevatron \cite{pp}, in $B$ decays  \cite{Bdecay},
fixed target experiments\cite{fixed}, $\gamma p$ collisions 
\cite{photo}, $e^+ e^-$ annihilations at CLEO \cite{ee},  and
$Z^0$ decays at LEP \cite{keungz,choz,jungilz}. 
Polarized heavy quarkonium productions were also considered as an
independent check of the color-octet mechanism \cite{bra_chen,bene_roth}.
It is adopted in the calculations of the $\psi^{'}$ polarizations
at the Tevatron \cite{psipol}.
A new way to regularize the ultraviolet/infrared divergences in 
heavy quarkonium calculations was proposed in Ref.~\cite{chen}.
Also, some NRQCD  matrix elements relevant to $S$- and  $P$-wave
heavy quarkonium decays were calculated on the lattice \cite{hcdecay}. 
Some reviews of earlier literatures can be found  in Ref.~\cite{annual}.

In the color-singlet model, the prompt $J/\psi$ production rate in $Z^0$ 
decays is dominated by charm quark fragmentation \cite{zfrag}.
However, recent reports by OPAL collaboration \cite{opal} claim
that they have observed an excess of events for
$Z^{0} \rightarrow \Upsilon (nS) + X$(for $n=1,2,3$), 
larger than the theoretical expectation by a factor of $\sim 10$,
compared to the $b-$quark fragmentation contribution\cite{zfrag}.
Similar excess was also observed in the prompt $J/\psi$ and $\psi^{'}$
production in $Z^0$ decays, although the experimental errors are quite
large \cite{opal2}.  It turns out that the color-octet gluon fragmentation
suggested by Braaten and Fleming could fix this discrepancy through
$Z^0 \rightarrow q \bar{q} + g$ followed by color-octet gluon
fragmentation  into $J/\psi$ with emission of soft gluons
\cite{keungz,choz}.  
In Refs.~\cite{keungz,choz}, the energy distribution of the produced 
$J/\psi$ via the color-octet $c\overline c(^3S_1^{(8)})$ intermediate state
was shown to be dramatically different from that of the $J/\psi$ produced
via the color singlet mechanism.   Therefore, the $J/\psi$ energy
distribution  in the $Z^0$ decays could be another good test of the idea
of the color-octet mechanism. 
In Ref.~\cite{jungilz}, the present authors have considered the angular
distribution of $J/\psi$'s in $Z^0$ decays, and one can find whether the
color-octet  mechanism is working or not.

In this work, we suggest another observable,
the polarization of $J/\psi$ at LEP produced via the color-singlet and 
the color-octet mechanisms.  
In short, $J/\psi$'s produced  via the two channels,
$c\overline c(^3S_1^{(1)})\to J/\psi$ and 
$c\overline c(^3S_1^{(8)})\to J/\psi$,    
have distinctively different polarizations.
The $J/\psi$ produced by the  color-octet mechanism is  
about $19\%$ longitudinally polarized,
whereas $J/\psi$ by the  color-singlet mechanism
is about $29\%$ longitudinally polarized.

\begin{figure}
\vskip .5cm
\hbox to\textwidth{\hss\epsfig{file=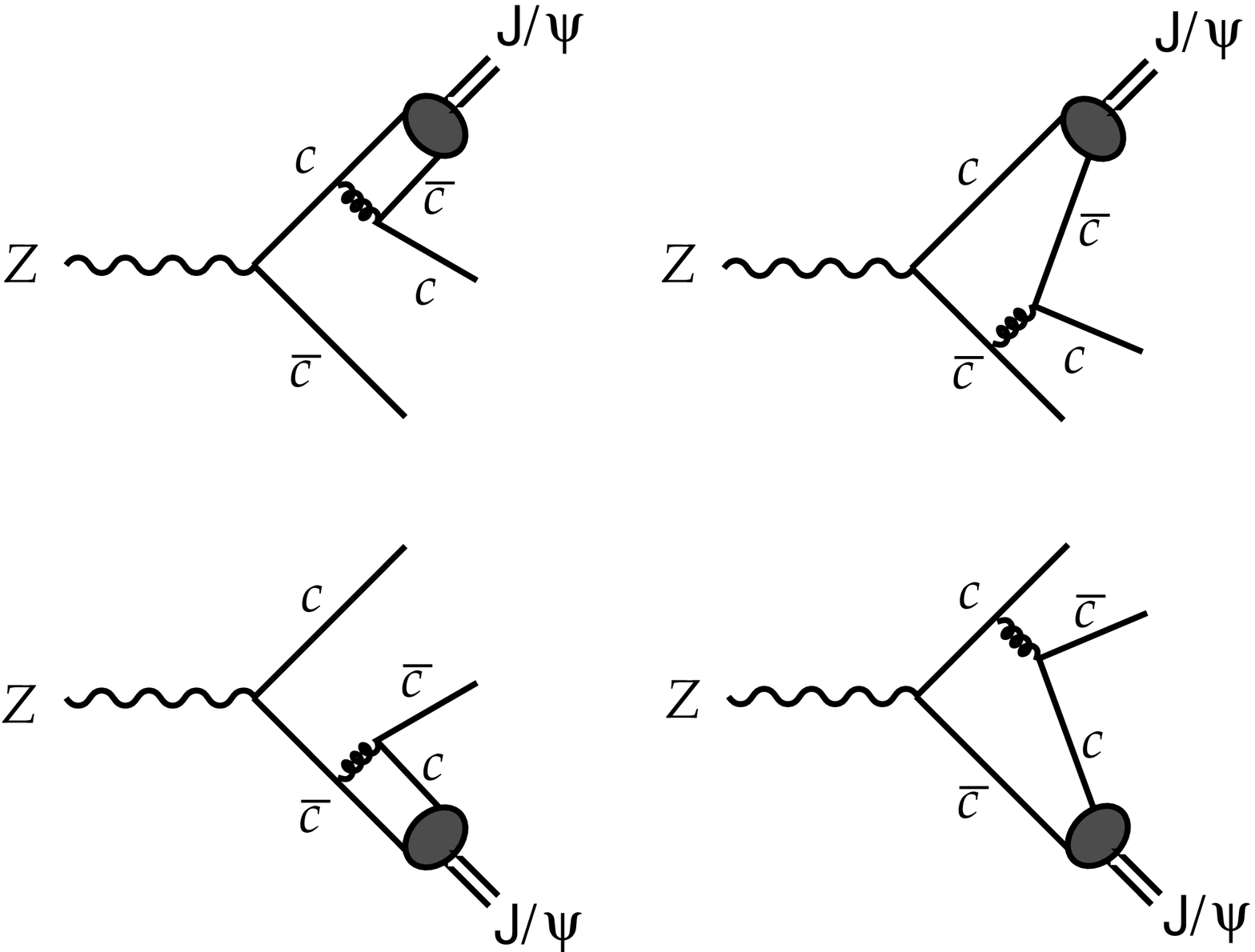,width=16cm}\hss}
\vskip 0.5cm
\caption{
}
\label{figone}
\end{figure}

\begin{figure}
\vskip .5cm
\hbox to\textwidth{\hss\epsfig{file=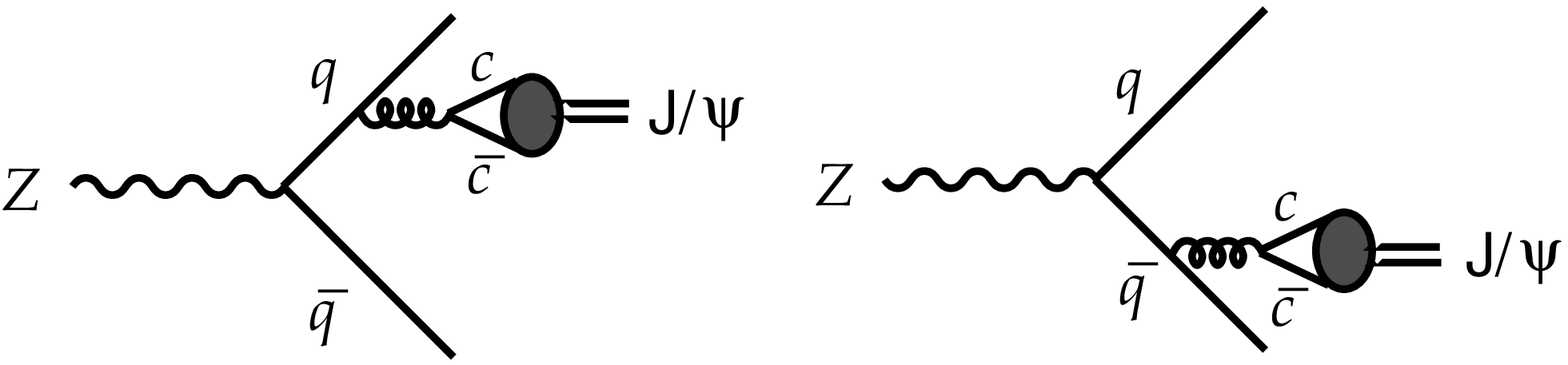,width=16cm}\hss}
\vskip 0.5cm
\caption{
}
\label{figtwo}
\end{figure}

When treating polarized quarkonium  productions, one should take care
of the soft process $Q\overline Q(^{2S+1}L_J)\to H$.
Recently, Braaten and Chen developed a method
for treating the polarized heavy quarkonium 
production\cite{bra_chen} which we use here. 
Beneke and Rothstein also pointed out that the 
interference among different $^3P_J$ states occurs in a polarized heavy 
quarkonium production~\cite{bene_roth}.
In general, one expresses the free particle amplitude, where 
$Q\overline Q$ makes a transition into the physical 
quarkonium state, in a power series of 
the relative momentum $\mbox{\boldmath$q$}$ of $Q$ and $\overline Q$ 
in the $Q\overline Q$ rest frame.
Then, one can find out what specific spectroscopic 
state of the $Q\overline Q$ pair is initially produced in the hard process
amplitude.
Finally, one may consider the soft transition in which the initially produced
$Q\overline Q$ system transforms into the  physical heavy quarkonium
state in which one is interested.
In the case of  heavy quarkonium production by the  color-singlet
mechanism, the soft process does not change any spectroscopic quantum
numbers such as color and angular momentum up to $v^2$ order correction
(which contains relativistic correction and double $E1$ transitions) :
\begin{eqnarray}
Q\overline Q(^{3}S_1^{(1)})\to Q\overline Q(^3P_J^{(8)})\to 
H(^{3}S_1^{(1)}).
\end{eqnarray}
In $Z^0$ decay, the color-singlet production process  
mainly comes from the Feynman diagram shown in Fig.~\ref{figone}, 
whereas  the color-octet production process mainly comes from
the soft process $Q\overline Q(^{3}S_1^{(8)})\to J/\psi$ 
as shown in Fig.~\ref{figtwo}.
If we consider $J/\psi$ polarization, we should 
take into account the relation of the $J/\psi$ polarization
and the  angular momentum of the initial $Q\overline Q$ pair.
If $J/\psi$ is produced via the color-singlet mechanism
\begin{eqnarray}
Q\overline Q(^{3}S_1^{(1)})\to J/\psi(^{3}S_1^{(1)}),
\end{eqnarray}
the polarization vector of $J/\psi$ is identical to spin wavefunction
of the initially produced $Q\overline Q$ pair.
The polarization vector of the $ J/\psi$ produced via the color-octet 
mechanism through double E1 transitions,
\begin{eqnarray}
Q\overline Q(^{3}S_1^{(8)})\to Q\overline Q(^{3}P_J^{(8)})\to
J/\psi(^{3}S_1^{(1)}),
\end{eqnarray}
is the same as the spin polarization vector 
of the initially produced $Q\overline{Q}(^{3}S_1^{(8)})$,
since the E1 transition conserves spin and the total angular momenta
of the $Q\overline{Q}(^{3}S_1^{(8)})$ and $J/\psi$ are the same.
Therefore, in these two channels, there is no problem, even though
we treat the  polarization vector of the produced $J/\psi$ and 
the spin wavefunction of the $Q\overline Q$ pair to be the same.
The only factor involving the polarization of the produced $J/\psi$
is the hard process shown in Figs.~\ref{figone} and \ref{figtwo}
which produce a $Q\overline Q$ pair at short distance. 
Since the $Q\overline Q (^{3}S_1^{(8)})$ produced via the color-octet mechanism 
comes from the gluon propagator, it seems to be strongly transversely
polarized. The quantitative number for the color-octet produced $J/\psi$ 
polarization can be obtained only after the full calculations, which
will be presented below along with the numerical results.

Before presenting the results for the $J/\psi$ polarization at LEP,
we first argue that the $Z^0$ polarization at LEP does not affect the 
$J/\psi$ polarization in $Z^0$ decays.  
The $Z^0$ produced at LEP is polarized as a result of unequal vector and 
axial vector couplings between electron and $Z^0$ boson. Therefore, the
density matrix $\rho_Z^{\mu\nu}$ of $Z^0$ is given by \cite{song86} 
\begin{equation}
\rho_Z^{\mu\nu} = \frac{1}{3} I^{\mu\nu}
                 -\frac{i}{2 M_Z} \varepsilon^{\mu\nu\lambda\tau} 
                          Z_\lambda {\cal P}_\tau
                 -\frac{1}{2} {\cal Q}^{\mu\nu},
\end{equation}
where $I^{\mu\nu} \equiv -g^{\mu\nu} +\frac{Z^\mu Z^\nu}{M_Z^2}$,
$Z^\mu$ is 4-momentum of $Z^0$. ${\cal P}^\mu$ and ${\cal Q}^{\mu\nu}$
are vector and tensor polarization of a $Z^0$ boson :
\begin{eqnarray}
{\cal P}^{\mu} & = & \frac{\Delta^\mu}{M_Z} ~\frac{g_V^2-g_A^2}{g_V^2+g_A^2}
\\
{\cal Q}^{\mu\nu} & = & -\frac{1}{3} I^{\mu\nu} 
              +\frac{\Delta^\mu\Delta^\nu}{M_Z^2},
\end{eqnarray}
where $\Delta^\mu \equiv (k_1 -k_2)^\mu$ with $k_1$ and $k_2$ being
four-momenta of $e^-$ and $e^+$  at LEP,
and $g_{V,A}$ are the vector and the
axial vector couplings between $e$ and $Z^0$ boson.
We can write the decay rate of $Z^0$ as 
\begin{equation}
d\Gamma = \frac{1}{2 M_Z} \rho_Z^{\mu\nu}
          \int [d p] \int d_2(PS) H_{\mu\nu},
\end{equation}
where $[d p] \equiv \frac{d^3 p}{(2 \pi)^3 2 p^0}$ is the invariant
phase space of  the produced heavy quarkonium with  four-momentum
$p_{\mu}$. By the Lorentz covariance, the integration $\int d_2(PS)
H_{\mu\nu}$ gives terms proportional
to $g_{\mu\nu}, Z_\mu Z_\nu, Z_\mu p_\nu, Z_\nu p_\mu, p_\mu p_\nu$,
and $\epsilon_{\mu\nu\alpha\beta} Z^\alpha p^\beta$.  
Here, $\epsilon_{\mu\nu\alpha\beta} Z^\alpha p^\beta$ is the only
non-vanishing term after being contracted  with the  
vector polarization term in $\rho_Z^{\mu\nu}$, 
and the result is proportional to $\cos \theta^*$, where $\theta^*$ is the 
angle between the initial electron-beam and the produced quarkonium 
directions.
When they are contracted with the tensor polarization contribution
in $\rho_Z^{\mu\nu}$, only $p_\mu p_\nu$ gives a nonzero quantity,
proportional to $3 \cos^2 \theta^* -1$.
In calculating the energy distribution or the total decay rate,
we integrate over the angle $\theta^*$ by  which 
all of the contributions from the polarization dependence vanish.
Therefore  $\rho_Z^{\mu\nu}$  can be safely replaced by 
$\frac{1}{3} I^{\mu\nu}$ in our calculations, effectively.

When one considers the $J/\psi$ polarization, it is convenient to 
define $\eta$ to be the ratio of the production rate ($\Gamma_L$) 
of the longitudinal $J/\psi$ to the total production rate
($\Gamma_{\rm TOT} \equiv \Gamma_{L} + \Gamma_{T}$)
as follows :
\begin{equation}
\eta \equiv \frac{\Gamma_L }{\Gamma_{\rm TOT} } = \frac{\Gamma_L}{
\Gamma_L + \Gamma_T }.
\end{equation}
This ratio $\eta$ can be determined experimentally from the measurement of 
the angular distribution of the leptons in the subsequent decay
$J/\psi \rightarrow l^+ l^-$ \cite{falk}.
Defining $\theta_{l}^{*}$ to be the angle between the three momentum of 
$J/\psi$ in the $Z^0$ rest frame and the three momentum of the 
daughter lepton (say $l^-$) 
in the rest frame of  $J/\psi$, the  angular  distribution  of a lepton
in the decaying $J/\psi$ rest frame has the form 
\begin{equation}
\frac{d\Gamma(J/\psi \rightarrow l^+ l^-)}{d \cos\theta_{l}^{*}}
   \propto 1 + \alpha \cos^2\theta_{l}^{*} ,
\end{equation}
where 
\begin{equation}
\alpha = \frac{1 - 3 \eta}{1 + \eta}.
\end{equation}
The unpolarized $J/\psi$ corresponds to $\eta = 1/3$, and $\alpha=0$.

The polarized $J/\psi$ production in $Z^0$ decays in the color-singlet model 
was calculated in Ref.~\cite{falk} using the fragmentation approximation.
In that paper, the authors showed that the asymmetry $\alpha$
is rather small, i.e. $\sim$ 5\% . 
Also, $\alpha$ is independent of the produced quarkonium mass so that
$\alpha$'s are the same both for $J/\psi$ and $\Upsilon$ production
in their fragmentation approach \cite{falk}.

In our work, we calculated all the Feynman diagrams without any fragmentation 
approximation in the color-singlet and color-octet contribution. 
We recover their
results in the limit of $\lambda \equiv m_{J/\psi(\Upsilon)}^2
/ M_Z^2 \rightarrow 0$.
Our results shown in the Appendix depend explicitly on the parameter
$\lambda$, and numerical values are shown in Table I.
Note that $\alpha$'s are considerably different for  $Z^0 \rightarrow
J/\psi + X$ and $Z^0 \rightarrow \Upsilon + X$.   Also, the 
fragmentation approximation is not that accurate at
calculating the $\Upsilon$ polarization in the $Z^0$ decays 
because of a rather large mass of 
$\Upsilon$.  $\alpha$'s are enhanced 
compared to those calculated in the fragmentation approximation.

When we compare the polarization of $J/\psi$ produced via the color-singlet
and the color-octet mechanisms, we observe that there is a considerable
difference between $\alpha_{1} = 0.10$ for the singlet and $\alpha_{8} =
0.36$ for the octet $c\overline c$ contribution to $J/\psi$ production.
Adding the singlet and the octet contributions, we get
$\alpha_{\rm tot}^{J/\psi} = 0.31$, which is appreciably different from 
$\alpha_{1}^{J/\psi} = 0.10$ or $\alpha_{\rm frag}^{J/\psi} = 0.053$.
These numerical values are consistent with those mentioned
in Ref.\cite{cheung2}. 
We have used the following numerical values for the matrix elements  of
NRQCD appearing in the $J/\psi$ production rates from the $Z^0$ decays :
\begin{eqnarray}
\langle 0 | {\cal O}_{1}^{J/\psi} ({^3S_1}) | 0 \rangle
& = &  0.73 ~{\rm GeV^3}
\\
\langle 0 | {\cal O}_{8}^{J/\psi} ({^3S_1}) | 0 \rangle
& = &  0.015 ~{\rm GeV^3}
\end{eqnarray}
We remark that both  $\alpha_{1}^{J/\psi}$ and $\alpha_{8}^{J/\psi}$
are independent of these NRQCD matrix elements,
since they cancel in the numerator and the
denominator when we take the ratio in $\eta$. On the other hand, 
$\alpha_{\rm tot}^{J/\psi}$ does depend on the numerical values of 
NRQCD matrix elements in Eqs.~(11)-(12),
each of which is known only within a factor of $\sim 2$.
Therefore, the definite test of the color-octet mechanism in 
$Z^{0} \rightarrow J/\psi + X$ will be a deviation
of the measured $\alpha_{\rm exp}^{J/\psi}$ from the singlet prediction,
$\alpha_{1}^{J/\psi}$ or $\alpha_{8}^{J/\psi}$, in the direction of a
larger value of $\alpha_{\rm exp}^{J/\psi}$.
Deviation of the $J/\psi$ polarizations (or $\alpha$) from the
color-singlet prediction ($\alpha_{1}^{J/\psi} = 0.10$)  may be
used as a probe to check the color octet mechanism in heavy quarkonium
productions, once a few thousand decays of $J/\psi \rightarrow l^+ l^-$ 
are observed in $Z^0$ decays.
In the case of $Z^{0} \rightarrow \Upsilon + X$, $\alpha^{\Upsilon}$ is 
not so sensitive to the singlet/octet mechanisms : 
$\alpha_8^\Upsilon = 0.28$ and $\alpha_1^\Upsilon = 0.23$, which are 
considerably larger than the prediction $\alpha_{\rm frag}^\Upsilon = 
0.053$ based on the fragmentation approach .

\begin{figure}
\vskip .5cm
\hbox to\textwidth{\hss\epsfig{file=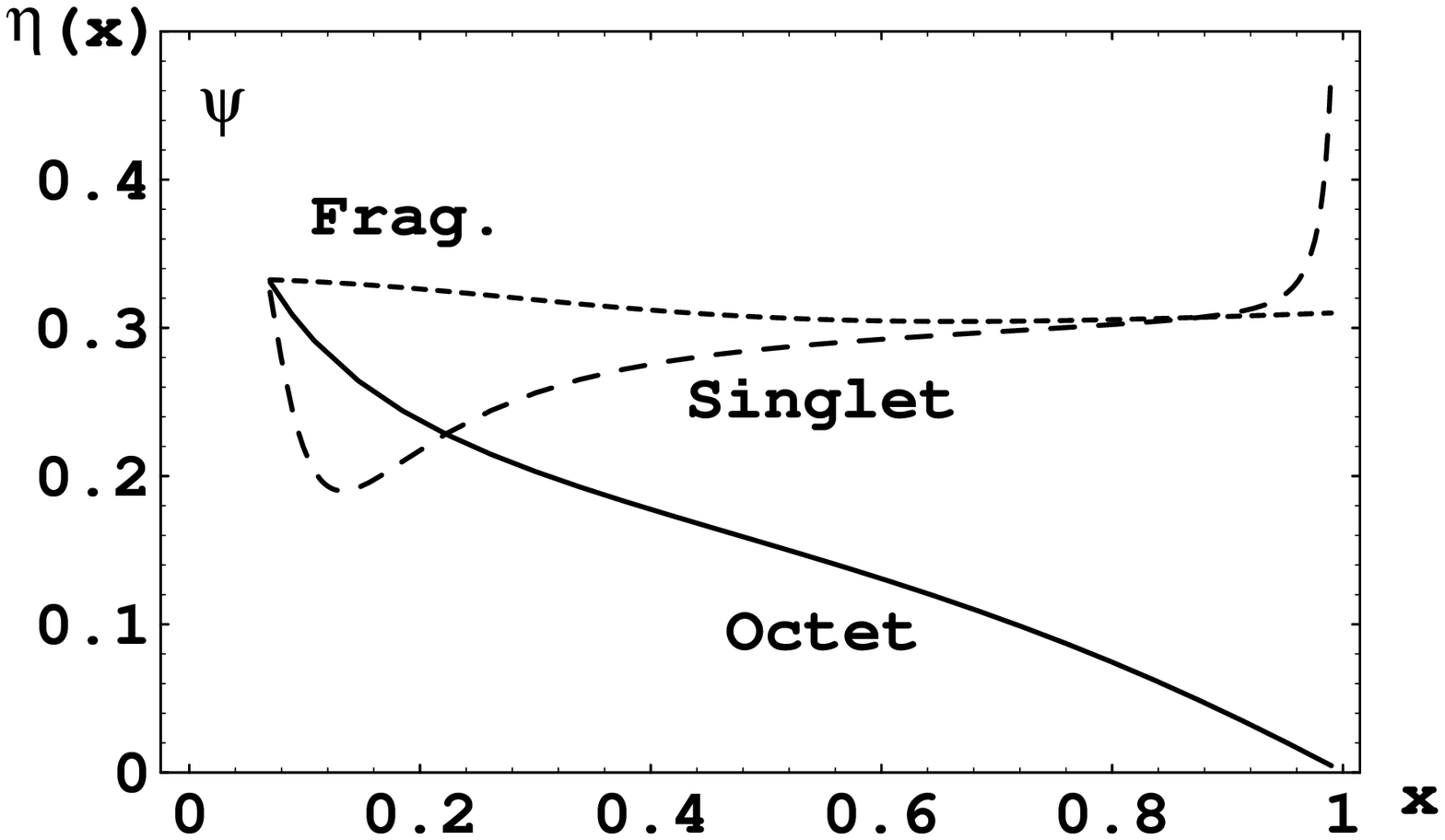,width=16cm}\hss}
\vskip 0.5cm
\caption{
}
\label{figthree}
\end{figure}

\begin{figure}
\vskip .5cm
\hbox to\textwidth{\hss\epsfig{file=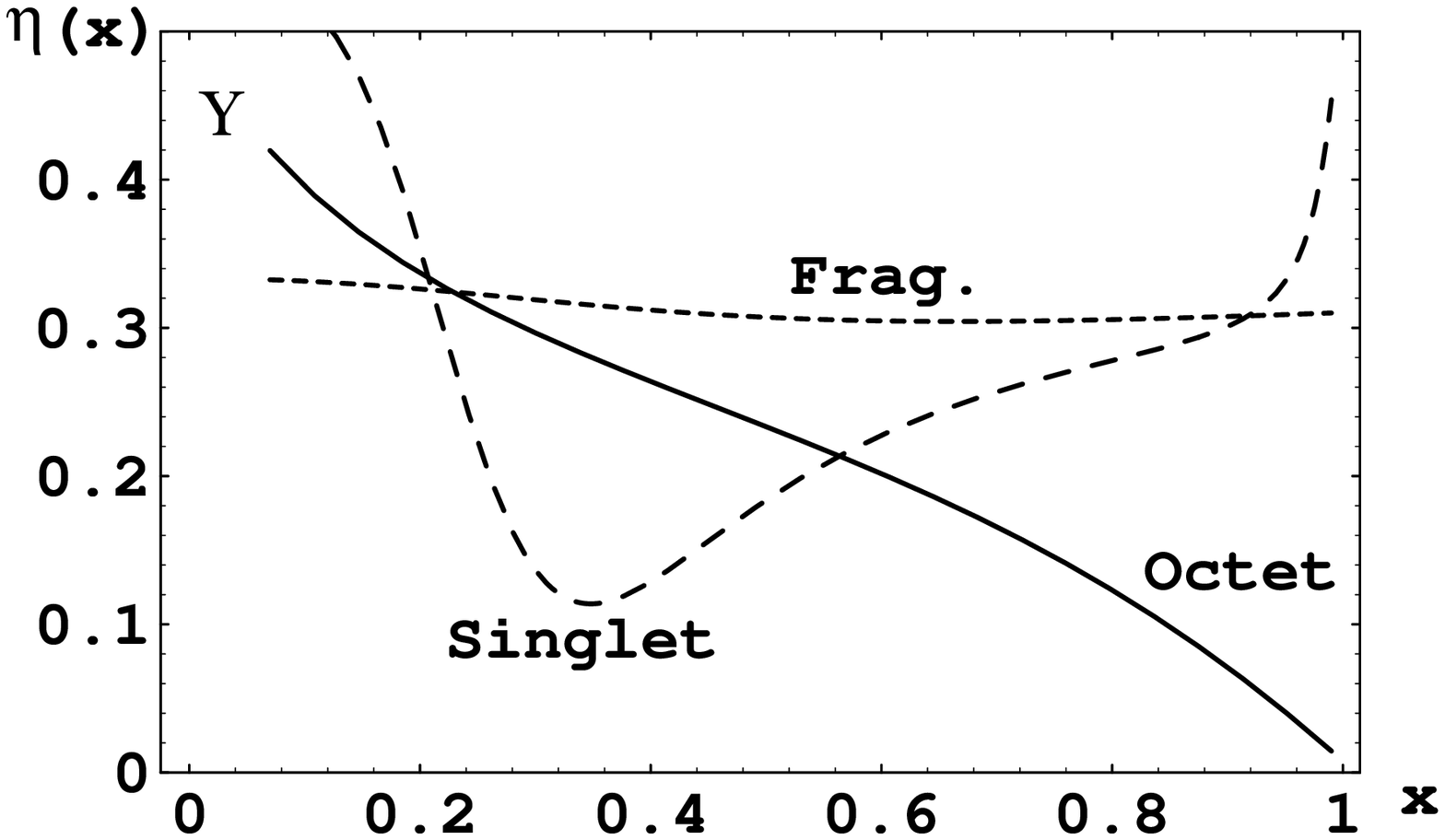,width=16cm}\hss}
\vskip 0.5cm
\caption{
}
\label{figfour}
\end{figure}

The energy dependence of $\eta(x)$ (with $x \equiv 2 E_{J/\psi} / M_Z$)
differs greatly depending on the $J/\psi$ production mechanisms,
as we can see in Fig.~\ref{figthree} for the case of $J/\psi$ and
Fig.~\ref{figfour} for the case of $\Upsilon$.
The $J/\psi$'s produced via the color-singlet mechanism are almost unpolarized
in almost the entire energy range, while the $J/\psi$'s produced 
via the color-octet mechanism are highly transversal (especially at high
energy).
Therefore if we observe quarkonium of a particular energy range, we can
greatly increase the polarization sensitivity if there are enough data.
For example, if we observe the $J/\psi$ only in the range of 
$0.7  \leq x \leq 0.9$, where most of color-singlet $J/\psi$ is
produced ~\cite{keungz} , $\eta_8^{J/\psi} = 0.076$ and $\eta_1^{J/\psi}
= 0.30$. These values correspond to $\alpha_8^{J/\psi} =0.72$ and
$\alpha_1^{J/\psi} =0.077$.
In the same energy range, $\eta_8^{\Upsilon} = 0.12$ and $\eta_1^{\Upsilon}
= 0.28$ for the case of $\Upsilon$, 
which correspond to $ \alpha_8^{\Upsilon}
= 0.57 $ and $ \alpha_1^{\Upsilon} = 0.13 $, respectively.
We can also observe $\Upsilon$'s in the energy range $0.3  \leq x \leq0.4$,
where most of color-octet $\Upsilon$'s are produced ~\cite{keungz}.
In this energy range, color-singlet $\Upsilon$ is more transversal
than color-octet $\Upsilon$, where $\eta_8^{\Upsilon} = 0.28$,
and $\eta_1^{\Upsilon} = 0.12$, corresponding to $\alpha_8^{\Upsilon}
= 0.13$ and $\alpha_1^{\Upsilon} = 0.57$ respectively.

In conclusion, we have calculated polarization  in 
heavy quarkonium ($J/\psi$ and $\Upsilon$) productions in $Z^0$ decay.
The polarization of $J/\psi$'s produced via the color octet mechanism
is more transversal compared to those produced via the color
singlet mechanism (Table~I).
Therefore, the measurement of polarizations provides
another independent test of  the idea of the color-octet mechanism.

\acknowledgements
We thank Prof. Dowon Kim for discussions on the experimental
situations. 
This work was supported in part by KOSEF through CTP at Seoul National 
University, in part by Korea Research  Foundations,
and in part by the Basic Science Research
Program, Ministry of Education,  Project No. BSRI--96--2418.
P.K. is also supported in part by NON DIRECTED RESEARCH FUND, 
Korea Research  Foundations.

\newpage
\section{appendix}

In this appendix, we show the analytic forms of $\Gamma_{1(8), L}$ and
$\Gamma_{1(8),{\rm TOT}}$, defined in Eq. ~(8).
 \def\la{\lambda}
\begin{eqnarray}
\Gamma_{8, {\rm TOT}}
&=& \frac{\alpha_s^2(2 m_c)}{18}
              \Gamma(Z \rightarrow q \overline q)
               \frac{\langle {\cal O}_8^{J/\psi}(^3 S _1)
   \rangle}{m_c^3}
\nonumber \\ 
& &\times \int_{2 \sqrt\la}^{1 +\la} d x 
 \left\{ \log\left(\frac{x +\sqrt{x^2 -4 \la}}{x -\sqrt{x^2 -4 \la}}\right) 
 \frac{[x^2 - 2 x +2 +2 \la (2 - x) + 2 \la ^2 ]}{x}
-2 \sqrt{x^2 -4 \la}  \right\},
\\
\nonumber \\
\Gamma_{8,L} &=& \frac{2 \alpha_s^2(2 m_c)}{9} 
            \Gamma(Z \rightarrow q \overline q)
            \frac{\langle{\cal O}_8^{J/\psi}(^3 S _1)\rangle}{m_c^3}
\nonumber \\ 
& &\times\int_{2 \sqrt{\la}}^{1 +\la} d x \frac{\la}{x^2 -4 \la} 
   \left\{ 
    \log\left(\frac{x +\sqrt{x^2 -4 \la}}{x -\sqrt{x^2 -4 \la}}\right) 
   \frac{[x - 1 + \la (x - 2) -  \la ^2 ]}{x} \right.
\nonumber \\
&  &\left. + \frac{1}{2 \la} (1 + \la) (1 +\la - x) \sqrt{x^2 -4 \la}
      \right\} .
\end{eqnarray}

\begin{eqnarray}
\Gamma_{1, {\rm TOT}}  &=& \frac{ \alpha_s^2(2 m_c)}{243}
          \Gamma(Z \rightarrow c \overline c)
        \frac{\langle{\cal O}_1^{J/\psi}(^3 S _1)\rangle}{m_c^3} 
\nonumber \\
& &\times\int_{2 \sqrt{\la}}^{1} d x  
{\Bigg[} 4 \la\log\left(\frac{x \sqrt{1 -x +\la} +\sqrt{(x^2 -4 \la) (1-x)}}
{x \sqrt{1 -x +\la} -\sqrt{(x^2 -4 \la) (1-x)}}\right){\bigg\{ }
\nonumber\\
& &\times 10 \la^3 (x^2 + 4)
 + \la^2 ( - 5 x^4 + 20 x^3 + 8 x^2 - 80 x + 80)\nonumber\\
& &+ \la (9 x^5 - 59 x^4 - 8 x^3 + 68 x^2 - 128 x + 64)
     + 4 x^2 (5 x^2 - 4)\nonumber\\
& &+\frac{g_V^2 -g_A^2}{g_V^2+g_A^2} {\Big(}
         2 \la^3 (x^2 + 4)
   + \la^2 (5 x^4 - 60 x^3 + 24 x^2 - 48 x + 48)\nonumber\\
& &+ \la x^2 ( - 9 x^3 + 73 x^2 - 76)
    + 32 x^4 ( - x + 1)  {\Big)}{\bigg\}} {\Big/}
\Big( x^3  (2-x)^2\Big) 
\nonumber\\
& & -8 \sqrt{\frac{(x^2 -4 \la) (1-x)}{1-x+\la}}{\bigg\{}
  2 \la^4 (x + 2) (5 x^3 - 38 x^2 + 60 x - 40)\nonumber\\
& &+ \la^3 ( - 5 x^6 + 66 x^5 - 286 x^4 + 888 x^3 - 992 x^2 + 960 x - 480)\nonumber\\
& &+ 6 \la^2 (2 x^7 - 25 x^6 + 118 x^5 - 324 x^4 + 384 x^3
- 360 x^2 + 288 x - 96)\nonumber\\
& &+ \la ( - 5 x^8 + 76 x^7 - 411 x^6 + 1168 x^5 - 1384 x^4
	       + 1248 x^3 - 1456 x^2 + 1024 x - 256)\nonumber\\
& &- 4 x^2 (x - 1)^2 (5 x^4 - 32 x^3 + 72 x^2 - 32 x + 16)\nonumber\\
& &+\frac{g_V^2 -g_A^2}{g_V^2+g_A^2} \Big(
   2 \la^4 (x + 2) (x^3 + 18 x^2 + 12 x - 8)\nonumber\\
& &+ \la^3 (5 x^6 - 78 x^5 + 274 x^4 - 840 x^3 + 384 x^2 + 448 x - 224)\nonumber\\
& &+ 4 \la^2 ( - 3 x^7 + 45 x^6 - 211 x^5 + 528 x^4 - 504 x^3
              + 44 x^2 + 144 x - 48)\nonumber\\
& &+ \la x^2 (x - 1) (5 x^5 - 73 x^4 + 328 x^3 - 712 x^2 + 560 x - 80)\Big)
\bigg\}    \Big/\Big( x^2  (2-x)^6\Big)\Bigg] 
\end{eqnarray}

\begin{eqnarray}
\Gamma_{1,L} &=& \frac{ \alpha_s^2(2 m_c)}{243}
          \Gamma(Z \rightarrow c \overline c)
        \frac{\langle{\cal O}_1^{J/\psi}(^3 S _1)\rangle}{m_c^3} \nonumber\\
& &\times\int_{2 \sqrt{\la}}^{1} d x  \Bigg[
   -4\la\log\left(\frac{x \sqrt{1 -x +\la} +\sqrt{(x^2 -4 \la) (1-x)}}
{x \sqrt{1 -x +\la} -\sqrt{(x^2 -4 \la) (1-x)}}\right)\bigg\{ \nonumber\\
& &  24 \la^4 (x^2 + 4)
   + 8 \la^3 (x^4 - 2 x^3 - x^2 - 24 x + 28)\nonumber\\
& &+ \la^2 ( - 3 x^6 + 24 x^5 - 128 x^4 + 64 x^3 - 112 x^2 - 128 x + 128)\nonumber\\
& &+ \la x^2 ( - x^5 + 3 x^4 + 56 x^3 + 60 x^2 - 64)
   + 4 x^4 ( - 5 x^2 + 4)\nonumber\\
& &+\frac{g_V^2 -g_A^2}{g_V^2+g_A^2}  \la \Big(
     8 \la^3 ( - x^2 - 4)
   + 8 \la^2 (4 x^4 - 12 x^3 + 7 x^2 + 12)\nonumber\\
& &+ \la x^2 (3 x^4 - 52 x^3 + 176 x^2 - 208 x + 16)
   + x^4 (x +2) (x^2 -3 x +6)\Big) \bigg\}\Big/
         \Big(x^3  (x^2-4 \la)  (2-x)^2\Big)
\nonumber\\
& &+8 \sqrt{\frac{(x^2 -4 \la) (1-x)}{1-x+\la}}\bigg\{
     8 \la^5 (x + 2) (9 x^3 - 62 x^2 + 108 x - 72)\nonumber\\
& &+ 8 \la^4 (3 x^6 - 25 x^5 + 76 x^4 + 176 x^3 - 584 x^2 + 912 x - 480)\nonumber\\
& &+ \la^3 ( - 9 x^8 + 108 x^7 - 756 x^6 + 2936 x^5
            - 8744 x^4 + 11424 x^3 - 12096 x^2 + 10752 x - 4224)\nonumber\\
& &+ 2 \la^2 ( - 15 x^8 + 216 x^7 - 1114 x^6 + 3520 x^5
                 - 3864 x^4 + 1984 x^3 - 2176 x^2 + 2304 x - 768)\nonumber\\
& &+ \la x^2 (3 x^8 - 36 x^7 + 149 x^6 - 128 x^5 - 1264 x^4
            + 2048 x^3 - 16 x^2 - 1280 x + 512)\nonumber\\
& &+ 4 x^4 (x - 1)^2 (3 x^4 - 24 x^3 + 64 x^2 - 32 x + 16)\nonumber\\
& &+\frac{g_V^2 -g_A^2}{g_V^2+g_A^2} 3 \la\Big(
   - 8 \la^4 (x + 2) (x^3 - 14 x^2 + 12 x - 8)\nonumber\\
& &+ 8 \la^3 (4 x^6 - 37 x^5 + 108 x^4 - 256 x^3 + 184 x^2 + 16 x - 32)\nonumber\\
& &+ \la^2 (3 x^8 - 88 x^7
   + 740 x^6 - 2600 x^5 + 5672 x^4 - 5728 x^3 + 1600 x^2 + 768 x - 384)\nonumber\\
& &+ 4 \la x^2 (4 x^6 - 53 x^5 + 231 x^4 - 624 x^3 + 844 x^2 - 496 x + 96)\nonumber\\
& &+ x^4 ( - x + 1) (x^5 - 13 x^4 + 56 x^3 - 128 x^2 + 80 x + 16)\Big)
\bigg\}\Big/\Big(3  x^2 (x^2 -4 \la) (2 - x)^6\Big) \Bigg],
\end{eqnarray}
with
\begin{equation}
\la \equiv \frac{4 m_c^2}{M_Z^2}  .
\end{equation}
The same formulas apply to the $\Upsilon$ case, with the substitution of
$m_b$ for $m_c$ , the corresponding change of couplings 
$g_V$ and $g_A$ , and the corresponding long-range matrix elements.


\begin{table}
\caption{
Longitudinal production fraction $\eta$ of quarkonium produced in the
$Z^0$ decay  and the asymmetry $\alpha$ of the angular distribution
of the quarkonium decay in its rest frame.
}
\label{table1}
\begin{tabular}{c|p{4.5cm}|p{4.5cm}}
 & $\eta^{J/\psi} / \eta^{\Upsilon}$ & $\alpha^{J/\psi} / \alpha^{\Upsilon}$
\\
\hline
Fragmentation ( Color Singlet) \cite{falk} & 0.31 /~~0.31 & 0.053 /~~0.053
\\
 Color Singlet (this work) & 0.29 /~~0.24 & 0.10 /~~0.23
\\
 Color Octet (this work) & 0.19 /~~0.22 & 0.36 /~~0.28
\\
 Octet +  Singlet (this work) & 0.21 /~~0.22 & 0.31 /~~0.28
\end{tabular}
\end{table}
\tighten
\newpage
\begin{center}
{FIGURE CAPTIONS}
\end{center}
\noindent
\vskip 1cm
Fig.1
\hskip .3cm
{
Feynman diagrams for  the color-singlet mechanism
for $Z^{0} \rightarrow (c \bar{c})({^{3}S_{1}^{(1)}}) + c \bar{c} $.
}
\\
\vskip 1cm
Fig.2
\hskip .3cm
{
Feynman diagrams for  the color-octet mechanism
for $Z^{0} \rightarrow q \bar{q} + J/\psi $ with  $q = u,d,c,s,b$.
}\\
\vskip 1cm
Fig.3
\hskip .3cm
{
Energy dependence of $\eta(x)$ in case of $Z^{0} \rightarrow J/\psi + X$ :
$\eta_8^{J/\psi}(x)$ in the solid curve,  
$\eta_1^{J/\psi}(x)$ in the dashed curve, and 
$\eta_{\rm frag}^{J/\psi}(x)$ in the dotted curve. 
}
\\
\vskip 1cm
Fig.4
\hskip .3cm
{
Energy dependence of  $\eta(x)$ in case of $Z^{0} \rightarrow \Upsilon + X$ :
$\eta_8^\Upsilon(x)$  in the solid curve, 
$\eta_1^\Upsilon(x)$  in the dashed curve,and
$\eta_{\rm frag}^{\Upsilon}(x)$ in the dotted curve. 
}
\\
\vskip 1cm

\end{document}